\documentclass{hep99}
\usepackage{epsf}
\newcommand{\AmS}{{\protect\the\textfont2
  A\kern-.1667em\lower.5ex\hbox{M}\kern-.125emS}}

\newcommand{\MeV}{\mbox{MeV}}
\newcommand{\fb}{f_{B{\scriptscriptstyle{d}}}}
\newcommand{\fbs}{f_{B{\scriptscriptstyle{s}}}}
\newcommand{\BBs}{B_{B{\scriptscriptstyle{s}}}}
\newcommand{\BB}{B_{B{\scriptscriptstyle{d}}}}
\newcommand{\fB}{f_{B{\scriptscriptstyle{d}}}}
\newcommand{\fBs}{f_{B{\scriptscriptstyle{s}}}}

\begin{document}

% declarations for front matter
\title{
\vspace{-2.5cm}
\begin{flushright}
{\normalsize
DAMTP-1999-117\\
UTCCP-P-74\\
%\vspace{-.2cm}
September 1999}
\end{flushright}
\vspace{.2cm}
Heavy-light decay constants from relativistic ${\bf N}_{\bf f}{\bf = 2,0}$ 
QCD%\footnote{talk presented by H.P.~Shanahan}
}

\author{
H.P.~Shanahan for the CP-PACS Collaboration
%CP-PACS Collaboration :
%      A.~Ali Khan\rlap,$^a$
%      S.~Aoki\rlap,$^b$
%      R.~Burkhalter\rlap,$^{\rm a,b}$
%      S.~Ejiri\rlap,$^{\rm a}$
%      M.~Fukugita\rlap,$^{\rm c}$
%      S.~Hashimoto\rlap,$^{\rm d}$
%      N.~Ishizuka\rlap,$^{\rm a,b}$
%      Y.~Iwasaki\rlap,$^{\rm a,b}$
%      K.~Kanaya\rlap,$^{\rm a,b}$
%      T.~Kaneko\rlap,$^{\rm a}$
%      Y.~Kuramashi\rlap,$^{\rm d}$
%      T.~Manke\rlap,$^{\rm a}$
%      K.~Nagai\rlap,$^{\rm a}$
%      M.~Okawa\rlap,$^{\rm d}$
%      H.~P.~Shanahan\rlap,$^{\rm e}$
%      A.~Ukawa\rlap,$^{\rm a,b}$ and
 %     T.~Yoshi\'e$^{\rm a,b}$ 
}

\address{%$^a$ Center for Computational Physics, 
%	University of Tsukuba, Tsukuba, Ibaraki 305-8577, Japan \\
%$^b$ Institute of Physics, University of Tsukuba, 
%	Tsukuba, Ibaraki 305-8571, Japan \\
%$^c$ Institute for Cosmic Ray Research,
%	University of Tokyo, Tanashi, Tokyo 188-8502, Japan \\
%$^d$ High Energy Accelerator Research Organization
%      	(KEK), Tsukuba, Ibaraki 305-0801, Japan \\
DAMTP, University of Cambridge, 
	Cambridge, CB3 9EW, England, UK }

\abstract{
We present results on an analysis of the decay constants 
$\fB$ and $\fBs$ with two flavours of sea quark. The
calculation has been carried out on 3 different bare gauge couplings and 
4 sea quark masses at each gauge coupling, with $m_\pi/m_\rho$ ranging
from 0.8 to 0.6. We employ the Fermilab formalism to perform calculations 
with heavy quarks whose mass is in the range of the b-quark. 
A comparison with a quenched calculation using the same action 
is made to elucidate the effects due to the sea quarks.
}

% typeset front matter (including abstract)
\maketitle

\section{Introduction\footnote{CP-PACS Collaboration : 
      A.~Ali Khan, S.~Aoki, R.~Burkhalter, S.~Ejiri, M.~Fukugita, S.~Hashimoto,
      N.~Ishizuka,  Y.~Iwasaki, K.~Kanaya, T.~Kaneko, Y.~Kuramashi,  T.~Manke,
      K.~Nagai, M.~Okawa,  H.P.~Shanahan,  A.~Ukawa,  T.~Yoshi\'e, }}

\vspace{-5pt}
An accurate determination of the parameters $\fB \sqrt{\BB}$ and 
$\xi = \fBs \sqrt{\BBs}/  \fB \sqrt{\BB}$, in conjunction with the (future) 
experimental data on
$\Delta m_d$ (and  $\Delta m_s$) will provide excellent constraints on $|V_{td}|$ and
$|V_{ts}|/|V_{td}|$\cite{buras}. Their calculation in the quenched approximation, using 
the plaquette gluon action, has been carried out using a  number of different 
formulations for heavy quarks and the results are converging\cite{shoji}.

A major uncertainty in these results, however, is that they may be susceptible to 
large corrections due to 
the effect of sea quarks \cite{booth,sharpe}. The penultimate step in eliminating this 
uncertainty is to consider the effect of two flavours of sea quark ($N_f=2$). 
Here we present the status of such a study, using the clover action for heavy quark, 
which we started last year\cite{hugh}.

As the lattice spacings available to us are relatively coarse, we deal with the 
large mass of $b$ quark $m_b a > 1$ in the formalism of Ref.~\cite{flabi}, which 
has been previously applied in the quenched calculation of B meson decay 
constants\cite{flabii,jlqcd}.  

In order to reduce discretisation effects, we have employed an RG-improved 
action in the gluon sector.  
Since this action was not considered in previous 
studies of $f_B$, we repeat the calculation in quenched QCD to 
compare with full QCD results. 

A comparison is also made with preliminary NRQCD results for the decay constants 
obtained on the same full QCD configurations\cite{arifa}.

\begin{figure}[tb]
%\vspace{-30pt}
\centerline{\epsfysize=5.4cm \epsfbox{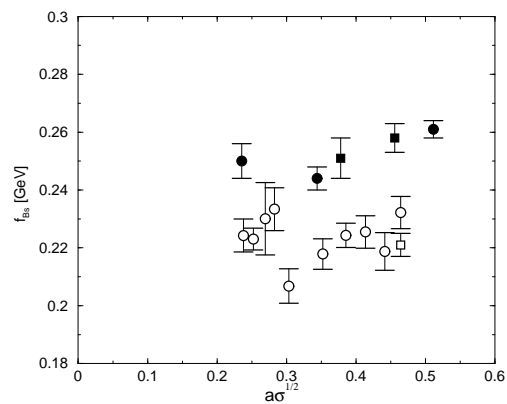}}
%\vspace{-30pt}
\caption{A comparison of $\fbs$ for $N_f=2$  (closed symbols) and quenched 
(open symbols) QCD. Circles indicate the relativistic action and squares indicate NRQCD.   
The spatial extent  of the lattices is roughly 2.6 fm.
The scale is set by string tension $\sqrt{\sigma}=427 \, \MeV$.
}
\label{fig:fbs}
%\vspace{-20pt}
\end{figure}

\begin{figure}[tb]
%%\vspace{-30pt}
\centerline{\epsfysize=5.4cm \epsfbox{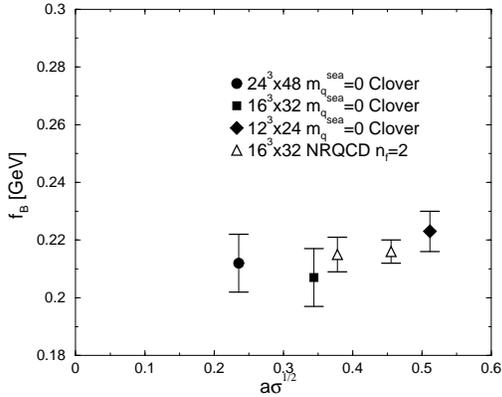}}
%\vspace{-30pt}
\caption{A comparison of preliminary NRQCD and clover results for $\fb$.}
\label{fig:fb}
%\vspace{-20pt}
\end{figure}

\begin{figure}[tb]
%\vspace{-30pt}
\centerline{\epsfysize=5.4cm \epsfbox{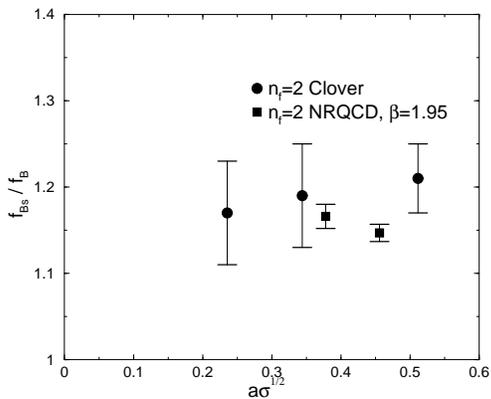}}
%\vspace{-30pt}
\caption{The $\fbs/\fb$ as a function of $a^2 \sigma$ for the relativistic and NRQCD actions.}
\label{fig:fbs_on_fb}
%\vspace{-14pt}
\end{figure}

\vspace{-5pt}
\section{Results}
\vspace{-5pt}
The computational details of this calculation are described in \cite{fbrel}.
In Fig.~\ref{fig:fbs} we plot results for $\fbs$ for $N_f=2$ 
and in the quenched approximation ($N_f=0$); 
sea quark mass is extrapolated to the chiral 
limit for $N_f=2$. A clear increase of 10--20\% is seen from two flavours of sea quark. 

We also include the preliminary results from NRQCD \cite{arifa} in Fig.~\ref{fig:fbs}, 
for two values of finite sea quark mass at $\beta=1.95$ in full QCD (filled squares), 
and one value in quenched QCD (open squares).  
In Fig.~\ref{fig:fb} we present a similar 
comparison of the Fermilab and NRQCD approaches for $\fb$. In both figures we find
good consistency of results between the two approaches. 
 
Finally, we plot the ratio $\fbs/\fb$ in Fig.~\ref{fig:fbs_on_fb}
using the $K$ meson mass to set the strange quark mass.
We observe only mild variation of the ratio with respect to $a$ for even the 
coarsest of our lattice spacings. 

As a preliminary result we quote 
\begin{eqnarray}
\fBs^{n_f=2} &=& 251\pm 3  \pm 4 (m_s)  \pm^{15}_{27} (\mbox{fit})\; \MeV, \\
 f_B^{n_f=2} &=& 210\pm 7  \pm^6_{14} (\mbox{fit})\; \MeV, \\
\frac{\fBs^{n_f=2}}{f_B^{n_f=2}} &=& 1.20\pm 4  \pm 2 (m_s) 
\pm^3_4 (\mbox{fit}) \; .
\end{eqnarray}
The first errors are statistical. The error labelled $(m_s)$ is due to the ambiguity 
of using the mass of the $\phi$ or $K$ to set the 
strange quark mass.
The central values are determined by assuming the results  independent of 
$a$  for the two finer lattice spacings.  The resulting 
systematic error $(\mbox{fit})$
is derived by taking the difference between a constant and a linear fit in
$a$ for all three points.
We should also add an uncertainty due to the choice of scale; our preliminary 
estimate is 15--20 MeV from comparision of results using the scale determind from 
$m_\rho$.

\vspace{-5pt}
\section{Conclusions}
\vspace{-5pt}
At this point it seems clear that there exists a systematic difference between the 
$N_f=0$ and $2$ data. Encouragingly enough, the preliminary NRQCD results are
also in agreement with the relativistic results in both cases as well.
One worrying point is that the quenched results for $\fbs$ is approximately 10\% larger
than the quenched results using the Wilson action\cite{shoji}. 
Clearly this effect needs further examination. 
The ratio $\fbs/\fb$ appears to be less affected by discretisation 
effects and is not substantially different from previous quenched 
calculations.  It should be noted that even a systematic  error of 10\%
for this ratio would be of substantial use for phenomenologists. 
It seems plausible then that a calculation of $\fbs/\fb$  could be
carried out for three flavors of dynamical quarks on a comparatively coarse lattice.
%assuming that the scaling violations will not be large.

%\vspace{10pt}

This work is supported in part by the Grants-in-Aid
of Ministry of Education,
Science and Culture (Nos.~09304029, 10640246, 10640248, 10740107,
11640250, 11640294, 11740162).
%SE and KN are JSPS Research Fellows.
HPS is supported by Research for the Future
Program of JSPS, and also by the Leverhulme foundation.


\vspace{-5pt}
\begin{thebibliography}{9}
\vspace{-5pt}
\bibitem{buras}For a recent review, see {\it e.g.,} 
A.J.~Buras, proceedings of the Lake Louise 
Winter Institute, Feb. 14-20, 1999, TUM-HEP-349/99, hep-ph/9905437.

%\bibitem{draper} T.~Draper, Nucl. Phys. Proc. Suppl. {\bf 73} (1999) 43.

\bibitem{shoji} S.~Hashimoto
to appear in the Proceedings of Lattice '99, 28 June-3 July 1999, Pisa, Italy.


\bibitem{booth} M.J.~Booth, Phys. Rev. {\bf D}51, 2338, (1995).

\bibitem{sharpe} S.R.~Sharpe and Y.~Zhang, Phys. Rev. {\bf D}53, 5125 (1996).

\bibitem{hugh}
H.~P.~Shanahan {\it et al.}, CP-PACS Collaboration,
Nucl. Phys. Proc. Suppl. {\bf 73} (1999) 375.

\bibitem{flabi}  A. X. El-Khadra, A. S. Kronfeld and P. B. Mackenzie,
Phys. Rev. {\bf D55} (1997) 3933.

\bibitem{flabii} A.X.~El-Khadra, {\it et al.},
Phys. Rev. {\bf D58} (1998) 014506.

\bibitem{jlqcd} S.~Aoki {\it et al.} (JLQCD Collaboration),
Phys. Rev. Lett. {\bf 80} (1998) 5711.

\bibitem{arifa}
A.~ Ali Khan {\it et al.}, CP-PACS collaboration,
to appear in the Proceedings of Lattice '99.

\bibitem{fbrel}
H.P.~Shanahan  {\it et al.} , CP-PACS collaboration,
hep-lat/9909052. 



%\bibitem{ruedi}
%R.~Burkhalter, Nucl. Phys. Proc. Suppl. {\bf 73} (1999) 3.


%\bibitem{bernard}
% C.W.~Bernard, J.N.~Labrenz \& A.~Soni Phys. Rev. {\bf D}49,
%2536, (1994).
\end{thebibliography}
\end{document}